\documentclass[aip, reprint, amsmath,amssymb,citeautoscript]{revtex4-1}
\usepackage{graphicx}
\usepackage{dsfont}
\usepackage{xcolor}
\draft 

\begin{document}


\title{Reducing network size and improving prediction stability of reservoir computing} 




\author{Alexander Haluszczynski}
\email{alexander.haluszczynski@gmail.com}
\affiliation{Ludwig-Maximilians-Universit\"at, Department of Physics, Schellingstra{\ss}e 4, 80799 Munich, Germany}
\affiliation{risklab GmbH, Seidlstra{\ss}e 24, 80335, Munich, Germany}
\author{Jonas Aumeier}
\email{jonas.aumeier@dlr.de}
\affiliation{Institut f{\"u}r Materialphysik im Weltraum, Deutsches Zentrum f{\"u}r Luft- und Raumfahrt,
M{\"u}nchner Str. 20, 82234 Wessling, Germany}
\author{Joschka Herteux}
\email{joschka.herteux@dlr.de}
\affiliation{Institut f{\"u}r Materialphysik im Weltraum, Deutsches Zentrum f{\"u}r Luft- und Raumfahrt,
M{\"u}nchner Str. 20, 82234 Wessling, Germany}
\author{Christoph R{\"a}th}
\email{christoph.raeth@dlr.de}
\affiliation{Institut f{\"u}r Materialphysik im Weltraum, Deutsches Zentrum f{\"u}r Luft- und Raumfahrt,
M{\"u}nchner Str. 20, 82234 Wessling, Germany}


\date{\today}

\begin{abstract}
Reservoir computing is a very promising approach for the prediction of complex nonlinear dynamical systems. Besides capturing the exact short-term trajectories of nonlinear systems it has also proved to reproduce its characteristic long-term properties very accurately. However, predictions do not always work equivalently well. It has been shown that both short- and long-term predictions vary significantly among different random realizations of the reservoir. In order to gain an understanding on when reservoir computing works best, we investigate some differential properties of the respective realization of the reservoir in a systematic way. We find that removing nodes that correspond to the largest weights in the output regression matrix reduces outliers and improves overall prediction quality. Moreover, this allows to effectively reduce the network size, and therefore increase computational efficiency. In addition, we use a nonlinear scaling factor in the hyperbolic tangent of the activation function. This adjusts the response of the activation function to the range of values of the input variables of the nodes. As a consequence, this reduces the number of outliers significantly and increases both the short and long-term prediction quality for the nonlinear systems investigated in this study. Our results demonstrate that a large optimization potential lies in the systematical refinement of the differential reservoir properties for a given data set.

\end{abstract}


\maketitle 

\begin{quotation}
A pervasive stigma of common machine learning methods is that they are considered an inscrutable black box. This is problematic for many practical applications, since a precise understanding of the tool is necessary to correctly assess uncertainties and sensitivities. Knowing that there is often significant variability in the prediction quality, the natural question arises how one can identify good predictions and prevent outliers that do not adequately resemble the targeted data or system. In contrast to many other neural network based approaches, reservoir computing makes it possible to bring light into the dark. Its comparably simple architecture allows for a systematic analysis of the differential properties of the reservoir realizations leading to good or bad predictions. In the context of nonlinear dynamical systems, a good prediction should not only be able to match the actual short-term trajectory but also needs to recreate its statistical long-term characteristics. To investigate the connection between properties of the reservoir and prediction quality, we remove certain nodes from the reservoir network and analyze how this impacts predictions. We find that a controlled node removal of the "right" nodes not only leads to less variability, and thus better predictions, but also allows to reduce network size noticeably. Furthermore, we turn from the reservoir itself to the activation function and show how rescaling of the argument gives rise to better results.

\end{quotation} 

\section{Introduction}
The remarkable rise of machine learning (ML) techniques during the recent years has made it inevitable for researchers to dig deeper into the mechanisms and properties of the methods. This is required to fundamentally understand how, when and why they are working well. Otherwise, the application of machine learning techniques to various fields in business and science carries the risk of misinterpreting the results if deeper methodological knowledge is lacking. 

In the context of complex systems research, the use of reservoir computing (RC) \cite{jaeger04}  -- also know as \textit{Echo State Networks} \cite{jaeger2001echo, maass2002real} -- for quantifying and predicting the spatiotemporal dynamics of nonlinear systems has attracted much attention recently \cite{lu17, pathak2017using,pathak18a,pathak18b, zimmermann18,carroll18,lu2018attractor,antonik18}.
RC represents a special kind of recurrent neural networks (RNN). The core of the model is a neural network called reservoir, which is a complex network with loops. Input data is fed into the nodes of the reservoir, which are connected according to a predefined network topology (mostly random networks). Only the weights of the linear output layer, transforming the reservoir response to output variables, are subject to optimization via linear regression. This makes the learning extremely fast, comparatively transparent and omits the vanishing gradient problem of other RNN training methods. The reservoir is kept fix and only the weights constituting the output layer are optimized in a deterministic and non-iterative manner. Therefore, RC allows for a controlled differential manipulation of the properties of the neural network and to identify, how those changes are associated with the prediction quality. \\ 

Many of the achievements obtained with RC --  be it e.g. the cross-prediction of variables in two-dimensional excitable media \cite{zimmermann18}, the reproduction of the spectrum of Lyapunov exponents in lower dimensional (Lorenz or R{\"o}ssler) and higher dimensional (Kuramoto-Sivashinsky) systems \cite{lu17, pathak2017using,pathak18a} or the prediction of extreme events \cite{doan2019physics} -- are impressive and significantly extend the possibilities to predict future states of high dimensional, nonlinear systems.
While the results reported in the works mentioned above are mainly based on a single or few realizations of reservoir computing, we showed, however, in an earlier study \cite{haluszczynski2019good} that there is a strong variability in prediction quality by running multiple realizations of the reservoir. The natural question that arises is where this variability comes from and whether one can associate good and bad predictions with differential properties of the reservoir. Based on a reservoir with unweighted edges, first attempts in this direction have been made by Caroll and Pecora \cite{carroll2019network}. They showed that symmetries in the network do have a considerable effect on the prediction quality of RC. In this work we investigate the effect of two methods to manipulate reservoirs with weighted edges, since those are typically used in time series prediction. First, we decrease the reservoir size by applying pruning techniques. Thinning out a (deep) neural network is a classical technique for enhancing its generalization ability. However, pruning mostly refers to the removal of synapses, i.e. edges, in a network. More rarely, pruning refers to the removal of neurons, i.e. nodes. 
So far, only few studies have investigated the effects of a controlled removal of edges or nodes in reservoir computing (see e.g. \cite{dutoit2009pruning, scardapane2014effective, scardapane2015significance}). Pruning of the reservoir network is a new optimization approach for the prediction of the long-term behavior of chaotic systems using RC. We propose and discuss a novel scheme for the controlled removal of nodes that relies on ideas stemming from network science. In addition, we vary the nonlinearity of the hyperbolic tangent activation function with a scaling factor. \\

The paper is organized as follows: Section~\ref{Methods} introduces reservoir computing and the reservoir manipulation methods used in our study. In section~\ref{Results} we present the main results obtained from the statistical analysis of the prediction results and its associated differential properties of the reservoir realizations. The summary and an outlook are given in section~\ref{Conclusion}.

\section{Methods}
\label{Methods}

\subsection{Reservoir Computing}
\label{Reservoir Computing}
\begin{figure}[b!] 
  \begin{center}
    \includegraphics[width=1.0\linewidth]{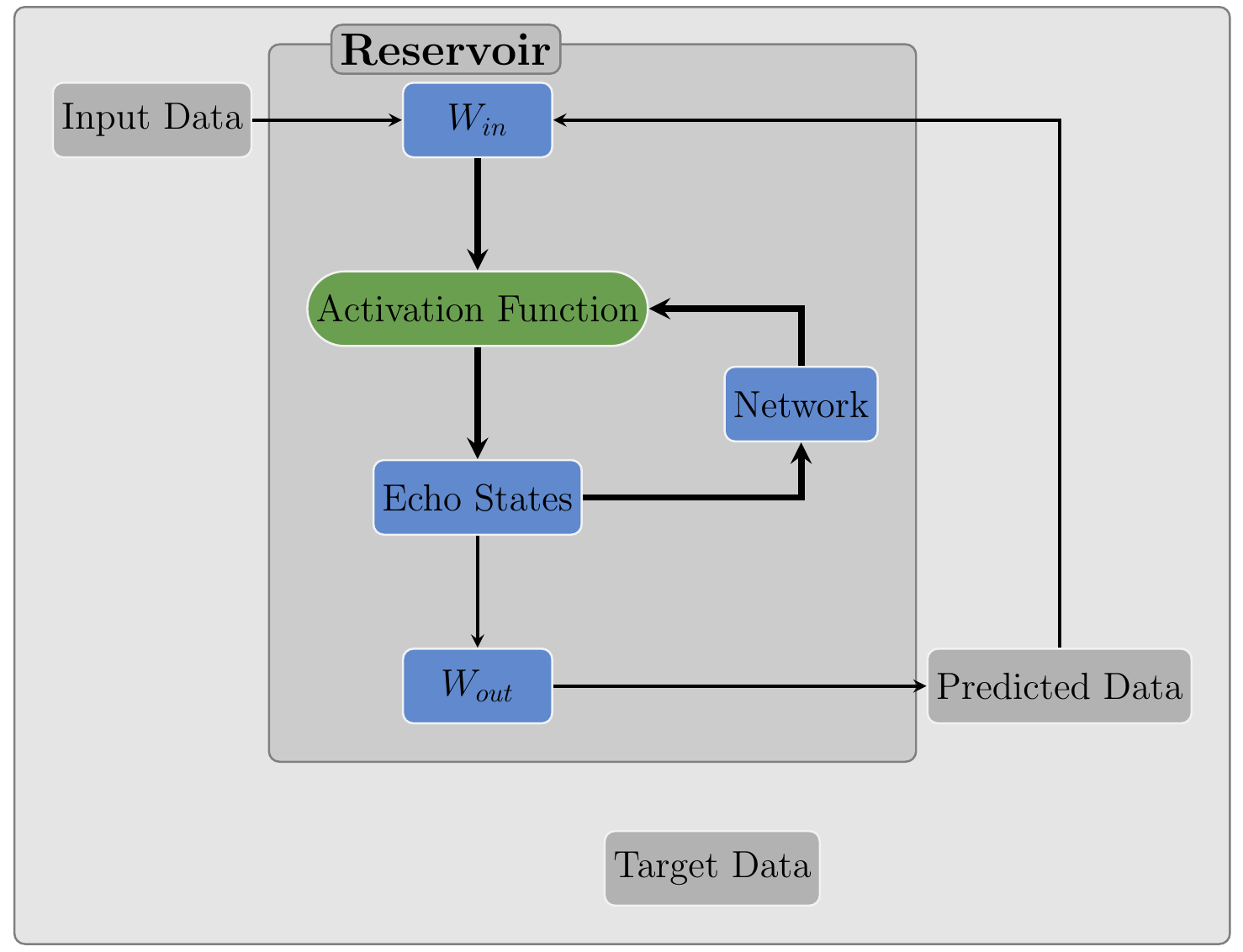}
    \caption{Schematic illustration of reservoir computing.}
    \label{fig:rescomp}
  \end{center}
\end{figure}

Within the class of artificial recurrent neural networks, reservoir computing is a promising approach for predicting complex nonlinear dynamical systems. The model is based on a static network called $reservoir$, whose nodes and edges are kept fixed after it has initially been set up. In contrast to feedforward type neural networks, the reservoir is allowed to have loops, and therefore past values are fed back into the system allowing for dynamics \cite{lukovsevivcius2009reservoir, genccay1997nonlinear}. As opposed to other neural network based machine learning approaches, the training process of reservoir computing alters only the linear output layer. This allows for large model dimensionality while still being computationally feasible \cite{lu2018attractor}.

This implementation is mainly based on the setup of our previous study \cite{haluszczynski2019good} and works in the following way.  First we set up the reservoir $\textbf{A},$ which has dimensionality $D_{r}$ and is constructed as a sparse Erd{\"o}s-Renyi random network \cite{erdos1959random}. In our study we chose $D_{r}=200$ nodes that are connected with a probability of $p=0.02$. This results in an unweighted average degree of $d=4$ while the weights of the edges are determined by independently drawn and uniformly distributed numbers within the interval $[-1,1]$. Once created, the reservoir is fixed and does not change over time. The next task is to feed the $D$ dimensional input data $\textbf{u}(t)$ into the reservoir $\textbf{A}$. This requires an $D_{r} \times D$ input matrix $\textbf{W}_{in}$ that defines for every node the excitation by each dimension of the input signal. The entries of $\textbf{W}_{in}$ are chosen to be uniformly distributed random numbers within a certain range to be defined later.  

A key element of the system are its $D_{r} \times 1$ reservoir states $\textbf{r}(t)$, which represent the scalar states of the nodes of the reservoir. We initially set $r_{i}(t_{0})=0$ for all nodes and update the reservoir states in each time step according to the equation
\begin{eqnarray}
\scalebox{0.90}[1]{$\textbf{r}(t+ \Delta{t}) = \alpha \textbf{r}(t)  + (1 -\alpha) tanh(\textbf{A}\textbf{r}(t) + \textbf{W}_{in} \textbf{u}(t))$} \ .
\label{eq:updating}
\end{eqnarray} 
As in Pathak et al. \cite{pathak2017using}, we set $\alpha = 0$, and therefore do not mix the input function with past reservoir states. Now we have a dynamical system where the reservoir $\textbf{A}$ itself is static and its scalar states $\textbf{r}(t)$ change over time. 

The next step is to map the reservoir states $\textbf{r}(t)$ back to the $D$ dimensional output \textbf{v} through an output function $\textbf{W}_{out}$ 
\begin{eqnarray}
\textbf{v}(t + \Delta{t}) = \textbf{W}_{out}(\textbf{r}(t + \Delta{t}),  \textbf{P}) \ .
\label{eq:output}
\end{eqnarray} 
Here we assume that $\textbf{W}_{out}$ depends linearly on a matrix $\textbf{P}$ and reads $\textbf{W}_{out}(\textbf{r},\textbf{P}) = \textbf{P}\textbf{r}$. This means that the output of the system depends only on the reservoir states $\textbf{r}(t)$ and the output matrix $\textbf{P}$, which contains $D_r \times D $ degrees of freedom. Therefore, after acquiring a sufficient number of reservoir states $\textbf{r}(t)$, we have to choose $\textbf{P}$ such that the output $\textbf{v}$ of the reservoir is as close as possible to the known real data $\textbf{v}_{R}$. This process is called training. Specifically, the task is to find an output matrix \textbf{P} using Ridge regression, which minimizes 
\begin{eqnarray}
\sum^{}_{-T \leq t \leq 0} {\parallel  \textbf{W}_{out}(\textbf{r}(t), \textbf{P}) - \textbf{v}_{R}(t) \parallel}^2 - \beta {\parallel \textbf{P} \parallel}^2 \ ,
\label{eq:minimizing}
\end{eqnarray} 
where $\beta$ is the regularization constant. Setting $\beta>0$ prevents from overfitting by penalizing large values of the fitting parameters. The notation $\parallel \textbf{P} \parallel$ describes the sum of the square elements of the matrix $\textbf{P}$. For solving this problem, we are using the matrix form of the Ridge regression \cite{hoerl1970ridge}, which leads to
\begin{eqnarray}
\textbf{P} = {(\textbf{r}^{T} \textbf{r} + \beta \mathds{1})}^{-1} \textbf{r}^{T} \textbf{v}_{R} \ .
\label{eq:ridgematrix}
\end{eqnarray} 
The notion $\textbf{r}$ and $\textbf{v}_{R}$ without the time indexing $t$ denotes matrices, where the columns are the vectors $\textbf{r}(t)$ and $\textbf{v}_{R}(t)$ respectively in each time step. In our implementation, we chose $t_{train} = 5000$ training time steps while allowing for a washout or initialization phase of $t_{init} = 5000$. During this time, we do not "record" the reservoir states $\textbf{r}(t)$ in order to allow the system to sufficiently synchronize with the dynamics of the input signal. 

\begin{table}[b]
\label{tab:hyperopt2}
\begin{center}
\begin{tabular}{ l | l | l}
\hline
\hline
$\rho$ & 0.17 \\
\hline
$\textbf{W}_{in}$ scale &  0.17  \\
\hline
$\beta$  & $1.9 \times 10^{-11} $ \\
\hline
\hline
\end{tabular}
\end{center}
\caption{Results of the hyperparameter optimization}
\end{table}

Now that \textbf{P} is determined, we can feed the predicted state $\textbf{v}(t)$ back in as input instead of the actual data $\textbf{u}(t)$ by combining Eq~\ref{eq:updating} and Eq~\ref{eq:output}. This allows to create predicted trajectories of arbitrary length due to the recursive equation for the reservoir states $\textbf{r}(t)$:
\begin{eqnarray}
\begin{aligned}
\textbf{r}(t+ \Delta{t}) &= tanh(\textbf{A}\textbf{r}(t) + \textbf{W}_{in} \textbf{W}_{out}(\textbf{r}(t),\textbf{P})) \\
                                  &= tanh(\textbf{A}\textbf{r}(t) + \textbf{W}_{in} \textbf{P}\textbf{r}(t))  \ .
\label{eq:updatingprediction}
\end{aligned}
\end{eqnarray} 
An illustration of this reservoir computing framework is given in Fig~\ref{fig:rescomp}.

To find the most suitable parameter values for the spectral radius of the reservoir $\rho(\textbf{A})$, the scale for $\textbf{W}_{in}$ and the regularization constant $\beta$, we carried out a hyperparameter optimization. As reservoir computing system can be trained very quickly, we use a simple random search procedure with uniform sampling from the parameter space. \footnote{The search ranges for the hyperparameter optimization are : [0 to 2.5] for $\rho(\textbf{A})$, [0 to 2.0] for $\textbf{W}_{in}$ and [\( \log_e 10^{-11} \) to \( \log_e 0.1\)] for $\beta$, which has been scaled onto a logarithmic scale for better coverage of small values} The objective function is the forecast horizon, as defined in Section~\ref{Measures}, averaged over $N = 30$ realizations. The term \textit{realizations} means running reservoir computing with the exact same parameters but different random realizations of the reservoir \textbf{A} and the input function $\textbf{W}_{in}$. Each of the realizations is starting from randomly chosen coordinates obtained from simulating a very long trajectory of the Lorenz system. In order to assure independent trajectories, small scale uniform noise is added. The optimal values are shown in Table I. 

\subsection{Controlled Node Removal and Activation Function Adjustment}
\label{Modification}
\begin{figure}[b!] 
  \begin{center}
    \includegraphics[width=1.0\linewidth]{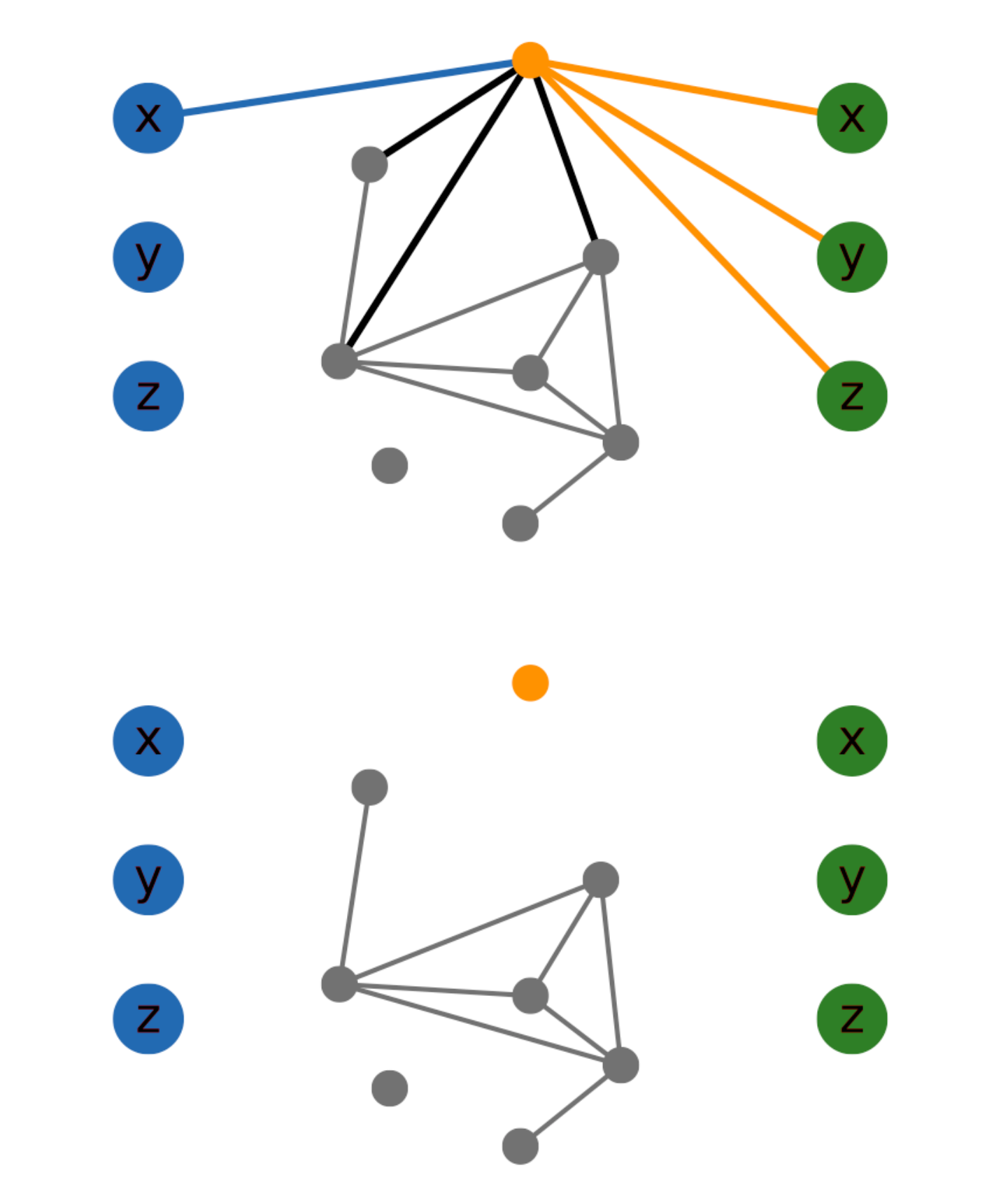}
    \caption{Schematic illustration of controlled node removal. The top graphic shows the initial network before the removal procedure. Here, the orange example node is fed only with input (blue) from the $x$ dimension of the system. The orange lines denote its contribution to the output values of all three dimensions. We assign the relevant weight by taking the maximum of the absolute value of these three weights. The black lines represent connections to other nodes. Input/output interactions of the other nodes are not shown here. In the bottom plot the example node has been removed, and therefore all connections and interactions vanished.} 
    \label{fig:nodeIllus}
  \end{center}
\end{figure}

The standard approach to reservoir computing exhibits a strong variability in prediction quality as shown in Haluszczynski et al. \cite{haluszczynski2019good}.
In order to reduce this variability, we make alterations to the reservoir structure by removing nodes and their respective edges from both the reservoir \textbf{A} and $\textbf{W}_{in}$. This is inspired by the concept of \textit{attack tolerance} \cite{albert2000error} in complex networks and the aim is to investigate the effect of removing nodes on the prediction capabilities of the system. The approach is motivated by the assumption that there is a relationship between the importance of each node and its output weights $\textbf{W}_{out}$ assigned in the training process. Following the findings of Albert et al. \cite{albert2000error} for networks, one would assume that the removal (“attack”) of important nodes (with high  $\textbf{W}_{out}$ values) has a large negative impact on the “response” of the reservoir to input data, i.e. on the prediction quality. On the other hand, the removal of unimportant nodes (with low $\textbf{W}_{out}$ values) should not alter the prediction too much. To test this assumption, we remove a fraction $p$ of the $N=200$ nodes, which correspond to certain values -- e.g. the largest or smallest -- of $\textbf{W}_{out}$. However, each node is affiliated not only with one but $D$ output weights, where $D$ denotes the dimensionality of the system that is being predicted. Hence, we sort $\textbf{W}_{out}$ based on the largest absolute value of all $D$ output weights for each node in order to determine which nodes should be removed. After removal, we train the newly obtained reduced network again. This leads to a new set of $\textbf{W}_{out}$. As a consequence, the new reservoir is not only reduced in size but also altered in its spectral radius, degree distribution and the distribution of $\textbf{W}_{in}$. The node removal process is illustrated in Fig~\ref{fig:nodeIllus}.

In addition to changes to the structure of the reservoir network outlined above, we study the effect of nonlinearity of the activation function. This has well-known effects on the memory of the reservoir \cite{inubushi2017reservoir,goudarzi2015exploring,verstraeten2007experimental,verzelli2019echo}. However, in the present study we focus on systems where the role of memory is small. To do this, we introduce a nonlinear scaling factor $a$ in the hyperbolic tangent of the activation function to further improve prediction quality. This changes the update equation for $\textbf{r}(t)$ to:
\begin{eqnarray}
\begin{aligned}
\textbf{r}(t+ \Delta{t}) = tanh(a [ \textbf{A}\textbf{r}(t) + \textbf{W}_{in} \textbf{P}\textbf{r}(t)])  \ .
\label{eq:scalingfac}
\end{aligned}
\end{eqnarray} 

The nonlinearity of the activation function is a crucial property for reservoir computing. Because both the reservoir itself and the output function are linear, the activation function is the only source of nonlinearity in the system. The introduction of a scaling factor in the argument can be interpreted as varying the degree of this nonlinearity. Equivalently, it can be seen as simply tuning the scale for $\textbf{W}_{in}$ and the spectral radius of $\textbf{A}$ simultaneously. Thus, the effective number of parameters stays the same. However, due to its relation to the activation function, it is interesting to study the isolated effect of the scaling, while fixing the other parameters.

\subsection{Measures and System Characteristics}
\label{Measures}
\subsubsection{Forecast Horizon} 
To quantify the quality and duration of the exact prediction of the trajectory, we use a fairly simple measure, which we call \textit{forecast horizon}. It is defined as the number of time steps wile the predicted and the actual trajectory are matching. As soon as one of the $D$ coordinates exceeds certain deviation thresholds, we consider the trajectories as not matching anymore. Throughout our study we use
\begin{eqnarray}
| \textbf{v}(t) - \textbf{v}_{R}(t) | > \boldsymbol{\delta} ,
\label{eq:fchor}
\end{eqnarray} 
where the thresholds are $\boldsymbol{\delta} = (5.8, 8.0, 6.9)^{T}$ for the Lorenz system. In general, the values are chosen based on the different ranges of the state variables and correspond to 15\% of the spatial extent of the attractor. The aim is that small fluctuations around the actual trajectory as well as minor detours do not exceed the threshold. Empirically we found that distances between the trajectories become much larger than the threshold values as soon as short-term prediction collapses. A similar measure has been proposed using a normalized L2 norm\cite{pathak18b}. However, when dealing with data, which shows significant differences in spatial extent between dimensions (e.g. the Chua circuit), this weighted approach is advantageous.

\subsubsection{Correlation Dimension} 
The structural complexity of a dynamical system is an important characteristic of its long-term properties. This can be quantified by its correlation dimension, where we measure the dimensionality of the space populated by the trajectory \cite{grassberger1983measuring}. The correlation dimension is based on the correlation integral 
\begin{eqnarray}
\begin{aligned}
C(r) &= \lim\limits_{N \rightarrow \infty}{\frac{1}{N^2}\sum^{N}_{i,j=1}\theta(r- | \textbf{x}_{i} - \textbf{x}_{j}  |)}      \\
       &= \int_{0}^{r} d^3 r^{\prime} c(\textbf{r}^{\prime}) \ ,
\label{eq:corrintegral}
\end{aligned}
\end{eqnarray} 
which describes the mean probability that two states in phase space are close to each other at different time steps. The condition \textit{close to} is met if the distance between the two states is less than the threshold distance $r$. $\theta$ represents the Heaviside function while $c(\textbf{r}^{\prime})$ denotes the standard correlation function. For self-similar strange attractors, the following power-law relationship holds in a range of $r$:
\begin{eqnarray}
C(r) \propto r^{\nu} \ .
\label{eq:corrdim}
\end{eqnarray} 
The \textit{correlation dimension} is then measured by the scaling exponent $\nu$. We use the Grassberger Procaccia algorithm \cite{grassberger83a} to calculate the correlation dimension of our trajectories. The scaling region is derived from the data itself as $r \in [0.5, 2.5]*s_{r}$, where the trajectory dependent scaling factor $ s_{r}$ is defined as $s_{r} =  \overline{\sigma(\textbf{u})}/8.5$. Thus, the scaling region depends on the standard deviation $\sigma$ of the input data $\textbf{u}$. This approach is purely data driven, and therefore does not require any knowledge about the system. 

\begin{figure*}[t!]
  \begin{center}
    \includegraphics[width=1\linewidth]{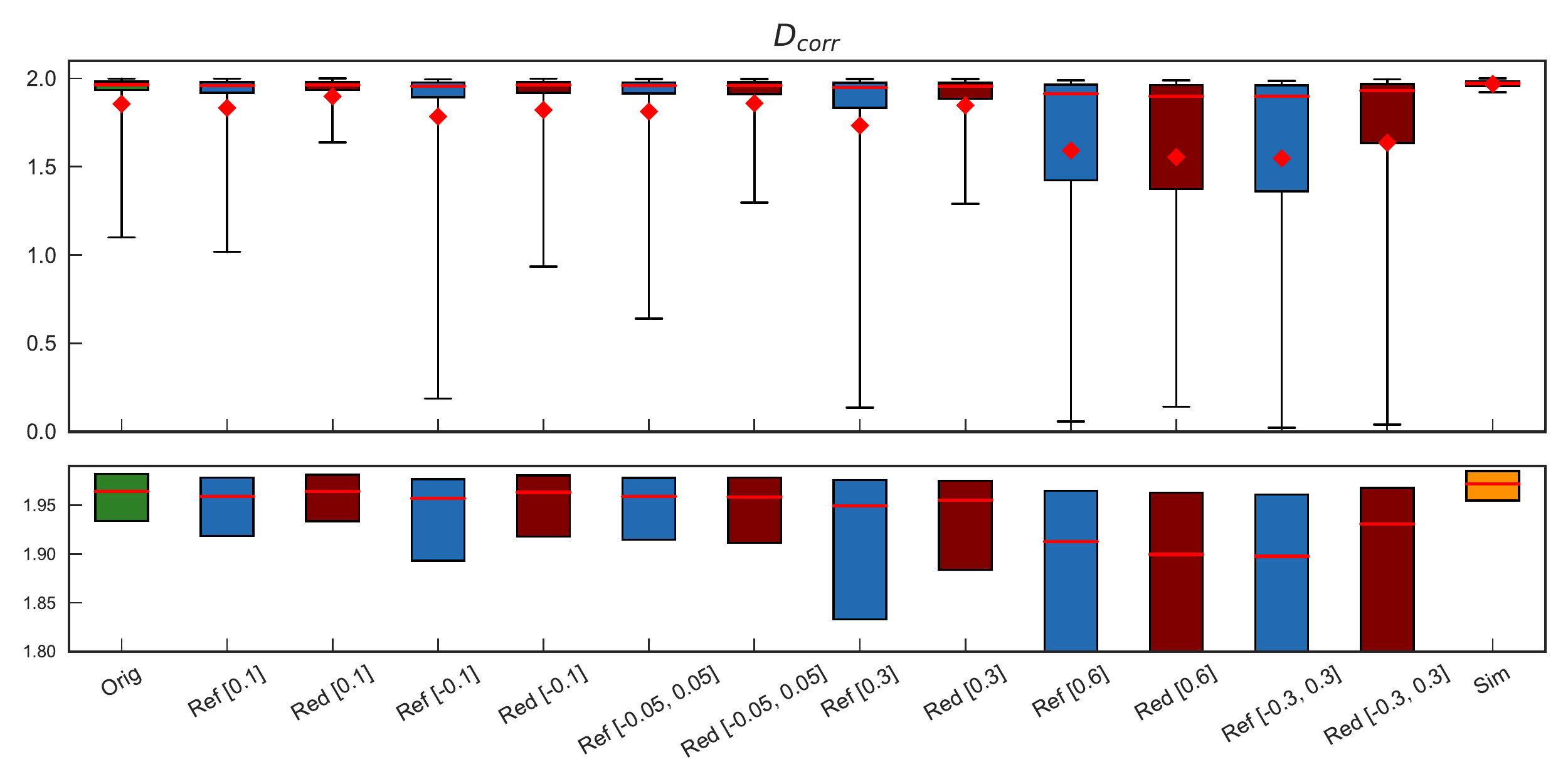}
    \caption{Boxplot of the correlation dimension for $N=500$ realizations for the original setup (green -- left) and different percentages of nodes removed. Positive numbers (e.g. Red [0.1]) represent a removal of the 10\% largest $\textbf{W}_{out}$ nodes, while negative numbers (e.g. Red [-0.1]) denote a removal of the 10\% smallest $\textbf{W}_{out}$ nodes. Consequently,  Red [-0.05,0.05] stands for the nodes with the 5\% largest and smallest $\textbf{W}_{out}$ values removed symetrically. \textit{Red} are the results for the system after node removal, while \textit{Ref} represents a smaller reference network. The yellow box on the right represents the error of the correlation dimension calculated from $N=500$ simulated trajectories. The boxes represent the 25\%--75\% percentile range while the extended lines denote the 5\% and 95\% percentile, respectively. Red bars are indicating the mean values and red dots show the median. In order to make a comparison easier, the bottom plot gives a zoomed in view of the 25\%--75\% percentile boxes and the respective median values.}
    \label{fig:removal1}
  \end{center}
\end{figure*}
\subsubsection{Largest Lyapunov Exponent} 
Apart from the structural properties, the temporal complexity of a system is another crucial feature of its so-called long-term climate. For chaotic systems, the analysis of the Lyapunov exponents is the most suitable approach for quantifying this. The Lyapunov exponents $\lambda_{i}$ describe the average rate of divergence of nearby points in phase space, and thus measure sensitivity to initial conditions. For each dimension in phase space there is one exponent. If the system exhibits at least one positive Lyapunov exponent, it is classified as chaotic, while the magnitude of the exponent quantifies the time scale on which the system becomes unpredictable \cite{wolf1985determining, shaw1981strange}. Therefore, it is sufficient for our analysis to determine only the largest Lyapunov exponent $\lambda_{1}$
\begin{eqnarray}
d(t) =  C e^{\lambda_{1} t} \ .
\label{eq:lyapunov}
\end{eqnarray} 
This makes the task computationally much easier than calculating the full Lyapunov spectrum. Here, $d(t)$ is the average distance or separation of the initially nearby states at time $t$ and $C$ is a constant that normalizes this initial separation. To calculate the largest Lyapunov exponent, we use the Rosenstein algorithm without requiring temporal separation of neighbors \cite{rosenstein1993practical}.


\subsection{Modified Lorenz system}
\label{Lorenz}
As an example for replicating chaotic attractors, we apply reservoir computing to the Lorenz system \cite{lorenz1963deterministic}. It has been developed as a simplified model for atmospheric convection and exhibits chaos for certain parameter ranges. The standard Lorenz system, however, is symmetric in $x$ and $y$ with respect to the transformation $x \rightarrow -x$ and $y \rightarrow -y$. In order to study a more general example, we would like to modify the Lorenz system such that this symmetry is broken. This can be achieved by adding the term $x$ to the $z$-component such that the equations read:
\begin{eqnarray}
\begin{aligned}
\dot x &= \sigma (y-x) \\
\dot y &= x (\rho-z)-y \\
\dot z &= x y - \beta z + x \ .
\label{eq:lorenz}
\end{aligned}
\end{eqnarray} 
This system is called modified Lorenz system. We utilize the standard parameters for its chaotic regime $\sigma = 10 , \beta = 8/3$ and $\rho = 28$ and solve the equations using the 4th order Runge-Kutta method with a time resolution $\Delta t = 0.02$. 

In addition to the Lorenz system, we run the analysis in section~\ref{Variability} also for a number of other nonlinear dynamical systems \cite{rossler1976equation, elwakil2002creation, chen1999yet, rabinovich1979stochastic, matsumoto1985double, thomas1999deterministic, rucklidge1992chaos} from the class of autonomous dissipative flows, such as the R{\"o}ssler system \cite{rossler1976equation}, Rabinovich-Fabrikant equations \cite{rabinovich1979stochastic}, Rucklidge system \cite{rucklidge1992chaos}, Halvorsen cyclically symmetric system \cite{sprott2003chaos} and the Chua circuit \cite{matsumoto1985double}. All these systems are $D=3$ dimensional but differ in properties like Lyapunov exponents, correlation dimension, size of the attractor and the nature of their nonlinearity. The parameters for all systems except Lorenz and R{\"o}ssler are taken from the textbook \textit{Chaos and Time-Series Analysis} by Sprott \cite{sprott2003chaos}.

\section{Results}
\label{Results}

In our previous study \cite{haluszczynski2019good} we showed that there is a strong variability in prediction quality by running the same setup with multiple different random realizations of the reservoir. In order to quantify the quality of a prediction, we analyzed both the exact short-term prediction horizon as well as the reproduction of the long-term climate of the system as measured by the correlation dimension and the largest Lyapunov exponent. Our aim is now to reduce this variability by applying the controlled node removal procedure and introducing an optimal choice for the nonlinear scaling parameter $a$ in the activation function as introduced in section~\ref{Modification}. 

\subsection{Controlled node removal}
\label{NodeRemoval}

After showing that changing the overall network topology e.g. by using small-world or scale-free networks does not lead to better predictions \cite{haluszczynski2019good}, we now focus on the differential properties of the reservoir. To do this, we carry out the controlled node removal procedure as introduced in section~\ref{Modification}. Here, we stick to the default setup by setting the nonlinear scaling parameter to $a=1$, and therefore do not rescale the activation function in order to separate effects. 

Figure~\ref{fig:removal1} shows a boxplot of the correlation dimension for $N=500$ realizations and compares the results for the original system (green box on the left) to those after different steps of node removal. In addition, the yellow box on the right shows the error of the correlation dimension calculated from $N=500$ simulated trajectories with different initial conditions. The boxes represent the 25\%--75\% percentile range while the extended lines denote the 5\% and 95\% percentile, respectively. Furthermore, the median values are indicated by the red bars while the red dots show the mean values. The labels on the x-axis are defined in the following way: \textit{Red [x]} denotes the results for the system after removing the nodes corresponding to the largest $x\%$ of the output weights if $x>0$ and smallest $x\%$ if $x<0$. Both positive and negative values at the same time mean that we symmetrically remove nodes from both ''sides''. In contrast, the results shown for the \textit{Ref [x]} labels are reference reservoirs, which are initially constructed and trained with less nodes and calibrated to the same spectral radius as the reservoirs after the node removal procedure. 

The results indicate that removing the nodes that correspond to the largest 10\% of the output weights -- \textit{Red [0.1]} -- improves the prediction quality compared to the original setup -- \textit{Orig}. In particular, the mean of the correlation dimension improves to $1.89$ compared to $1.85$ in the default setup, while the median stays at $1.96$. The values of the simulated system are $1.97$ and $1.97$. This means that predominantly bad predictions have been enhanced. Moreover, the 5\% percentile significantly increases from around 1 to 1.6. This indicates a lower number of outliers, where the reproduction of the correlation dimension did not work. As this reduced reservoir now effectively only has 180 nodes, it is interesting to analyze how a reservoir computing setup performs, which is initialized with only 180 nodes. We can see in Fig~\ref{fig:removal1} that for \textit{Ref [0.1]}, the reproduction of the correlation dimension becomes slightly worse as compared to the default setup with  $D_{r}=200$ nodes. This means that the improvement due to the controlled node removal is not due to the changed reservoir size but driven by the altered properties of the system. In contrast, removing nodes corresponding to the smallest 10\% of the output weights has a slightly negative effect on the prediction quality. However, the results are still better than those of its reference system with the same spectral radius and only 180 nodes initially. Finally, we symmetrically removed the nodes corresponding to the smallest and largest 5\% of the output weights. As for the first case, the prediction quality improves compared to the default system and the performance is again better than its reference system.
\begin{figure}[b]
  \begin{center}
    \includegraphics[width=1\linewidth]{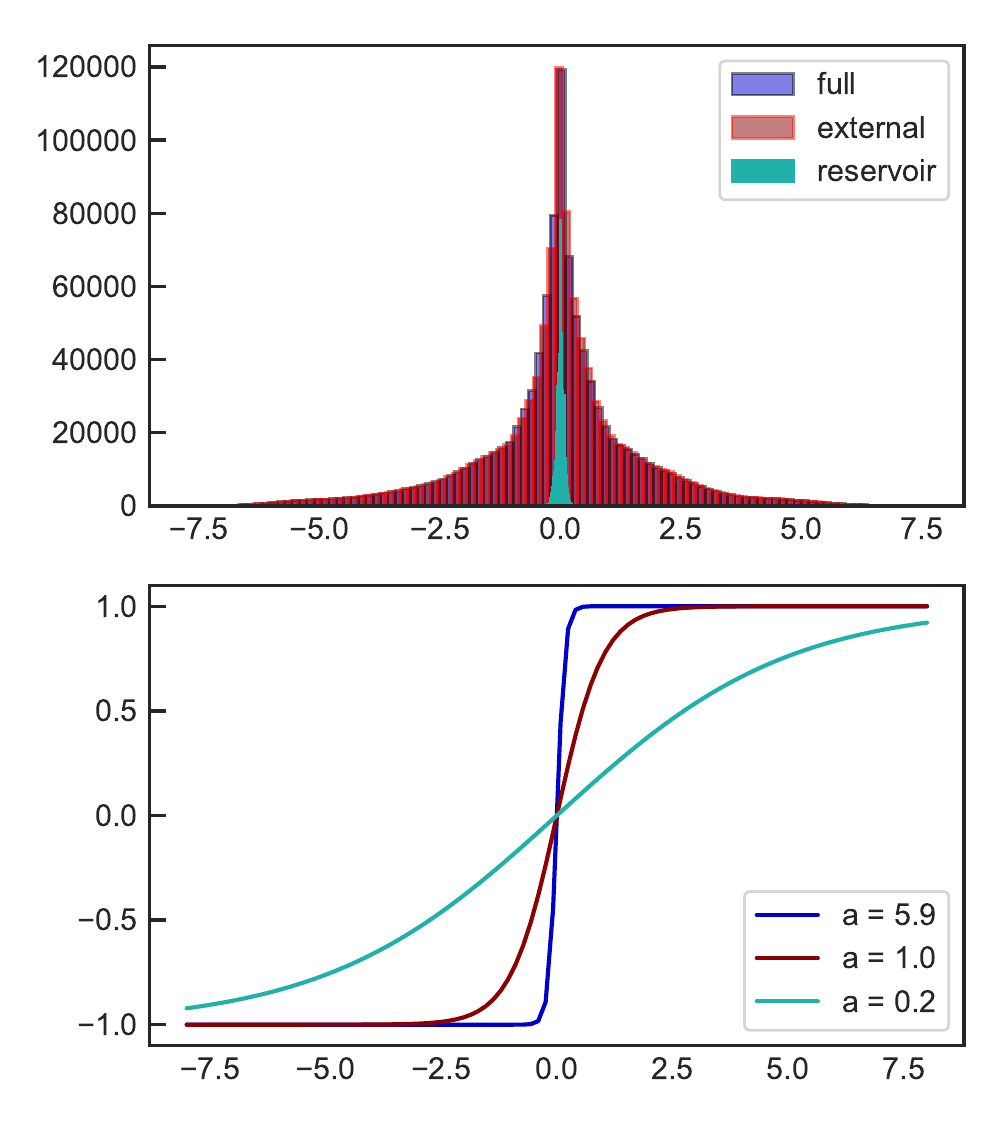}
    \caption{Top plot: Distribution of the arguments of the activation function during training period split into the contribution from the reservoir (green), the input term (red) and total (blue). Bottom plot: Hyperbolic tangent for different nonlinear scaling factors.}
    \label{fig:tanhRange}
  \end{center}
\end{figure}
\begin{figure*}[t!] 
  \begin{center}
    \includegraphics[width=1.0\linewidth]{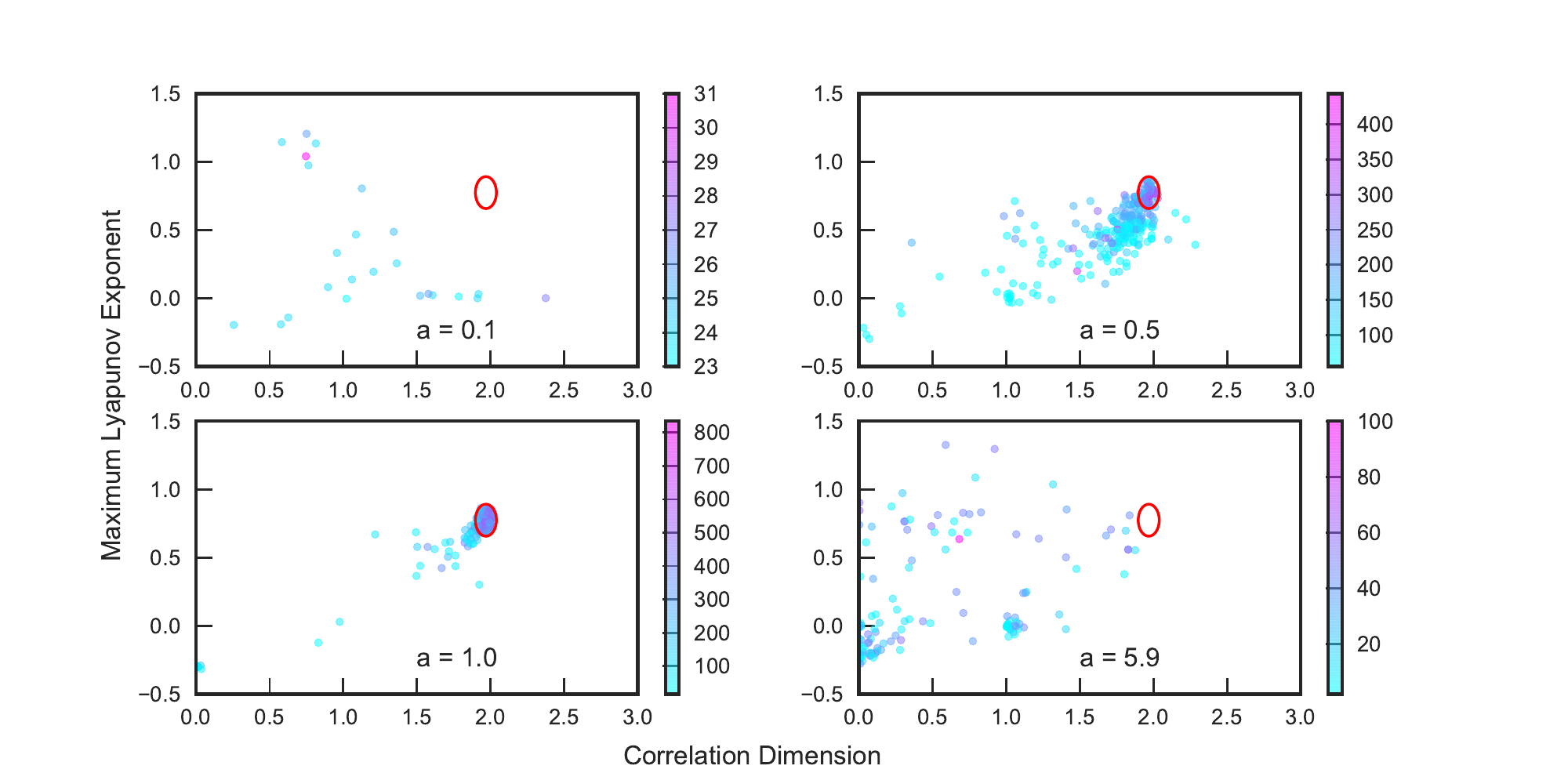}
    \caption{Largest Lyapunov exponent scattered against correlation dimension for different values of the nonlinear scaling parameter $a$ based on $N=300$ realizations each. The colors denote the forecast horizon of the predictions and the red ellipses show the three $\sigma$ errors of the correlation dimension ($\sigma = 0.024$) and the largest Lyapunov exponent ($\sigma = 0.039$) calculated from simulations of the actual system.}
    \label{fig:tanh}
  \end{center}
\end{figure*}

Naturally the question arises, how results change if we remove more than 10\% of the nodes and if it is possible to achieve comparable performance for smaller reservoir computing systems than the original setup with $D_{r}=200$ nodes. Thus, we calculated the results for removing up to 60\% of the nodes, and therefore significantly reduced network sizes. As a first step, we increase the percentage of removed nodes to those associated with the largest 30\% of the output weights. We can see that the performance is comparable to the larger original system while the number of outliers is still reduced. This can be observed by the shorter length of the black line. Moreover, this also holds for the mean and median values. Those are $1.85$ and $1.96$, respectively, for the reduced system and $1.86$ and $1.96$ for the original system. In addition, we also ran the same analysis for the nodes belonging to the smallest 30\% of output weights. Again, this leads to significantly worse results than excluding nodes with large weights. 

In contrast to the improvements in reproducing the correlation dimension seen for the 10\% and 30\% cases, removing the nodes corresponding to the largest 60\% of the output weights clearly leads to lower prediction quality and a higher number of bad realizations. The same can be observed for removing those nodes based on the smallest 60\% of the output weights, which is not shown here. It is interesting to note that in both cases the initially reduced reference system now performs better than in those cases, where a lower percentage of nodes has been removed. Furthermore, symmetrically removing the nodes reflecting both the largest and smallest 30\% of the output weights leads to better results than removing 60\% of either the largest or smallest. In addition, the results now outperform those of the reference system. While the overall prediction quality is notably worse than for the default system, it is very interesting to notice that it is possible to still achieve good prediction results with a significantly downscaled system. This can be very beneficial when applications are computationally more challenging, e.g. when the dimensionality of the dynamical system is high or the trajectories are very long. Moreover, reducing the network size is important when it comes to hardware implementations of reservoir computing, such as neuromorphic computing \cite{tanaka2019recent}. To make this more practicable, Griffith et al. \cite{griffith2019forecasting} proposed very simple reservoir designs with low connectivity. In contrast, our approach reduces the number of nodes $D_{r}$, and therefore could add additional benefits for hardware implementations.

Instead of calculating the correlation dimension, we ran the same analysis also based on the forecast horizon of the predictions. As the results look very similar to those of the correlation dimension, they are not shown here.

\subsection{Prediction variability and nonlinear scaling parameter}
\label{Variability}

As a next step, we focus on the activation function and examine the effect of different choices for the nonlinear scaling parameter $a$. The upper plot of Fig~\ref{fig:tanhRange} shows the distribution of the arguments of the hyperbolic tangent activation function during the training period. While the green area shows the influence of the reservoir term $\textbf{A}\textbf{r}(t)$, the red area represents the impact of the external input $\textbf{W}_{in} \textbf{u}(t)$. One can clearly see that the values of the reservoir term are very small compared to those of the external input. In commonly used parameterizations of reservoir computing, the value for the $\textbf{W}_{in}$ scale is 1 -- this means that the weights of the input function are uniformly distributed between -1 and 1. However, our hyperparameter optimization in section~\ref{Reservoir Computing} led to a $\textbf{W}_{in}$ scale of $0.17$, and therefore we can approximately say that the input scale of 1 in the standard parameterization is equivalent to a value of $a=5.9$ in our setup ignoring the comparably small influence of the reservoir term. If we compare the scale of the distribution to the functional form of the hyperbolic tangent in the lower plot, it becomes clear that for $a=5.9$ the majority of points lies in the saturation regime of the function. Intuitively one can expect that the best results can be achieved, if $a$ is chosen such that a large part of the distribution of the input arguments lies within the "dynamical" range of the hyperbolic tangent rather than in its saturation regime. Low values of $a$, however, would lead to an approximately linear behavior of the function, and would thus not allow the system to adequately capture the nonlinear dynamics of the input data.

In order to test this assumption empirically, we simulated $N=300$ realizations for different values of $a$. We then evaluated the forecast horizon as well as the long-term climate for each realization. The bottom right plot in Fig~\ref{fig:tanh} shows the largest Lyapunov exponent scattered against the correlation dimension for the modified Lorenz system. The results are based on the above described default setup with the nonlinear scaling factor set to $a=5.9$. The red ellipse shows the three $\sigma$ errors of the correlation dimension and the largest Lyapunov exponent. Those are calculated from simulations of the actual equations of the Lorenz system for $N=500$ different initial conditions. We can clearly see that for $a=5.9$ all points are widely spread outside this error ellipse, and are therefore to be classified as bad predictions. This is because they do not well resemble the long-term climate. While some realizations lead to meaningful values for the largest Lyapunov exponent, the correlation dimension is badly reconstructed in particular.

\begin{figure}[t!]
  \begin{center}
    \includegraphics[width=1\linewidth]{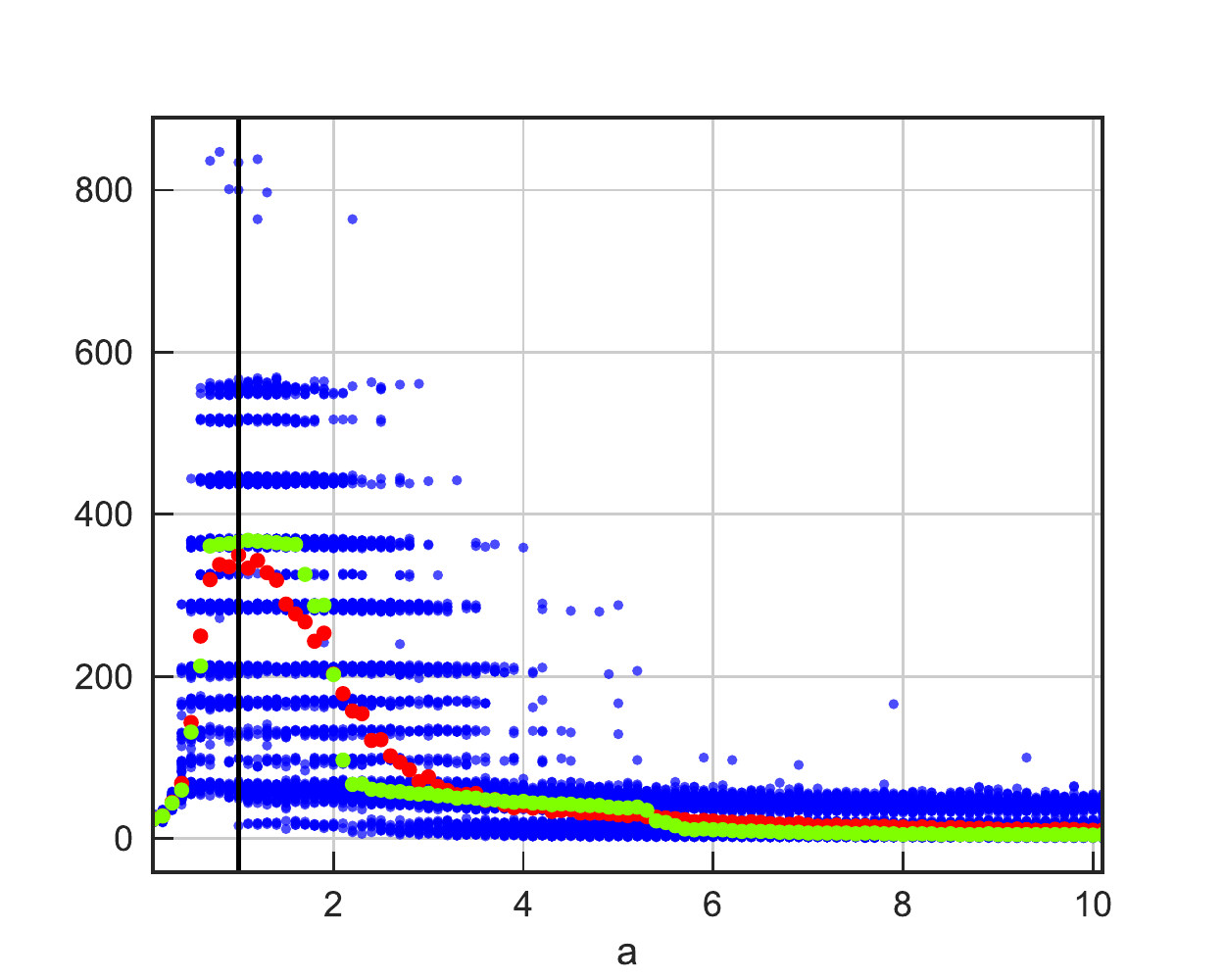}
    \caption{Forecast horizon of the modified Lorenz system plotted for different values of the nonlinear scaling parameter $a$ with $N=300$ realizations for each $a$ (blue). Furthermore, the average (red) as well as the median (green) value are shown for each value of $a$. The black horizontal line marks the optimal choice for $a$.}
    \label{fig:optimalA}
  \end{center}
\end{figure}

To find the optimal value for $a$, we systematically analyzed multiple realizations for a number of different values of $a$ between 0 and 10. This is shown in Fig~\ref{fig:optimalA}, where the blue points correspond to the forecast horizon of the single realizations. In addition, the red and green dots represent the average and median value across all realizations for a given value of $a$. We then determine the optimal value for $a$ such that the average is maximized. This leads to an optimal value of around $a=1.0$, which is in line with our expectation given that we carried out a hyperparameter optimization in the beginning. For validating the above arguments, we turn back to Fig~\ref{fig:tanh}. The bottom left plot shows the results for the optimal choice of $a$, where many outliers, and thus bad predictions disappeared. Moreover, there is now a compact cloud of points around the error ellipse, and therefore the overall prediction quality is significantly better as compared to the case $a=5.9$ in the bottom right plot. In contrast, setting $a=0.1$ and $a=0.5$ as shown in the top plots leads to a complete breakdown of the prediction ability of the system. The reason that one can see only a few points in the top left plot is the following. The prediction quality for $a=0.1$ completely collapses in most cases such that we obtain $NaN$ results for our calculations of the largest Lyapunov exponent. This happens in cases where the prediction jumps between multiple points in a cyclical fashion. Consequently, this leads to division by zero and generally only occurs for unsuitable parameter choices -- in this case for too small values of $a$. As both examples in the top plots correspond to arguments of the activation function being in the linear regime of the hyperbolic tangent, this demonstrates that nonlinearity in the activation function is essential for predicting complex nonlinear systems. Besides the results for the reproduction of the long-term climate, we also show the forecast horizon encoded in the colors of the points. Equivalently, the longest forecast horizon can be achieved by choosing the optimal value for $a$, whereas smaller or larger values both lead to worse results. Another interesting result is that realizations, which well resemble the long-term climate have a higher forecast horizon than those failing to properly reconstruct the climate.

\begin{figure}[t!] 
  \begin{center}
    \includegraphics[width=1\linewidth]{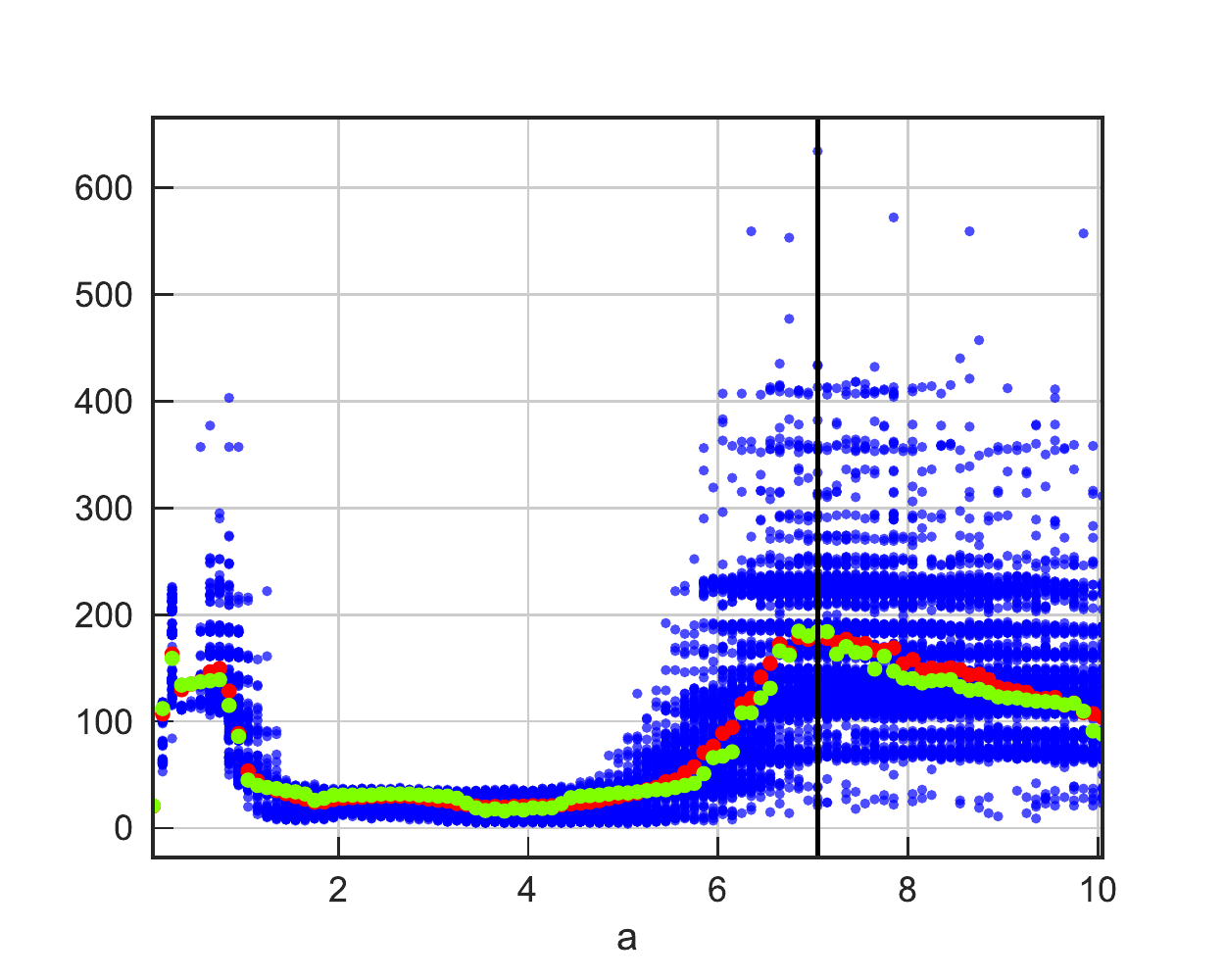}
    \caption{Forecast horizon of the Chua circuit plotted for different values of the nonlinear scaling parameter $a$ with $N=300$ realizations for each $a$ (blue). Furthermore, the average (red) as well as the median (green) value are shown for each value of $a$. The black horizontal line marks the optimal choice for $a$.}
    \label{fig:chua}
  \end{center}
\end{figure}

\begin{figure}
  \begin{center}
    \includegraphics[width=1\linewidth]{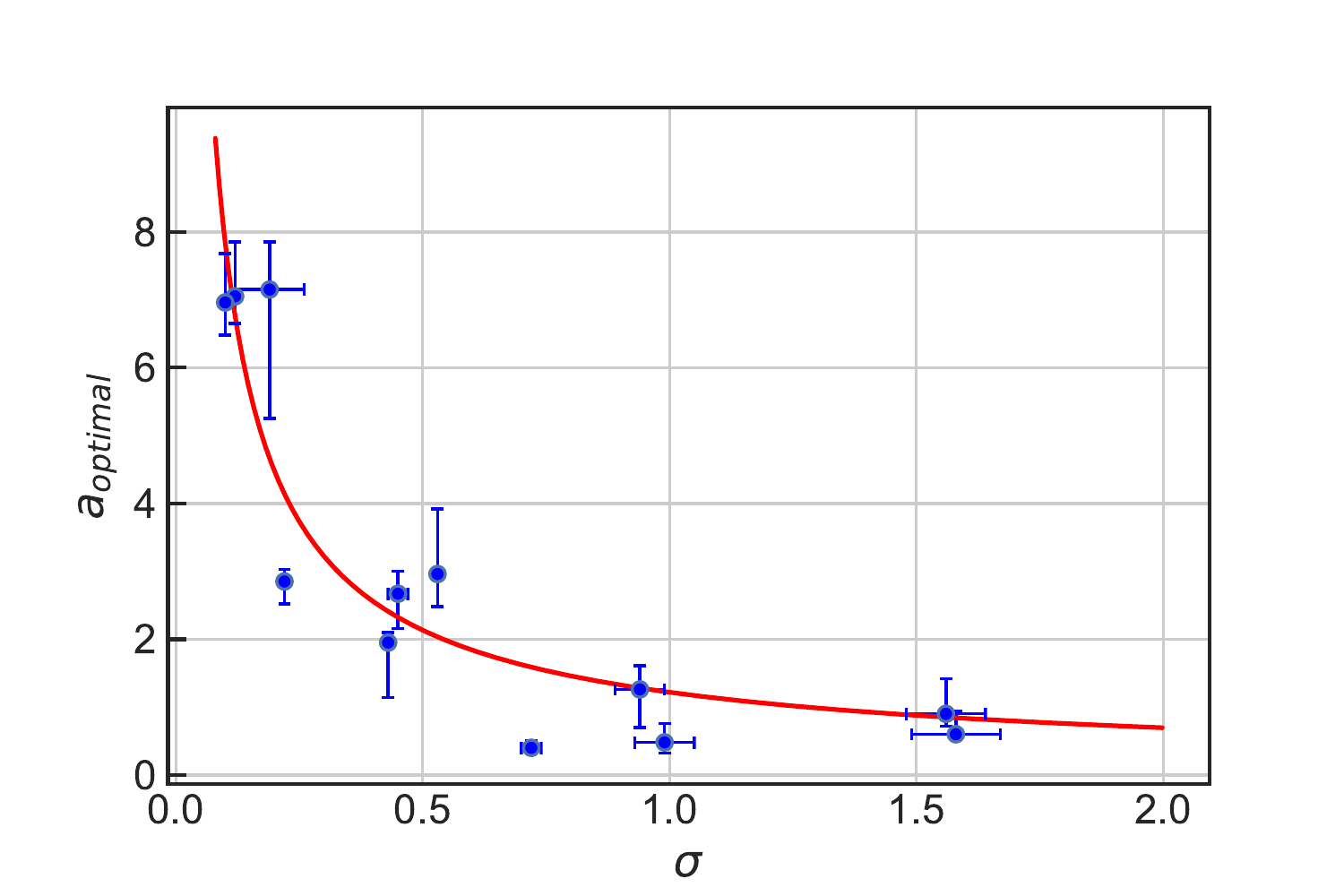}
    \caption{Blue dots: Optimal value for the nonlinear scaling parameter $a$ -- based on the maximum of the average forecast horizon over all realizations for a given $a$ -- plotted against the standard deviation $\sigma$ of the input data. The vertical bars denote the range of values for $a$, where the average forecast horizon is up to 10\% lower than for the optimal $a$, while the horizontal bars represent the standard error. The red line represents a fit through the points. Dynamical systems from left to right: Rabinovich-Fabrikant equations, Chua circuit, Complex Butterfly attractor, Thomas' cyclically symmetric attractor, R{\"o}ssler system, Rucklidge system, Halvorsen model, \textit{blended system:} 0.4*Modified Lorenz system + Halvorsen Model,  \textit{blended system:} 0.5*Modified Lorenz system + 3*Rabinovich-Fabrikant equations, \textit{blended system:} R{\"o}ssler + 2*Rucklidge system, Modified Lorenz system and Chen system}
    \label{fig:systems}
  \end{center}
\end{figure}

In addition to the Lorenz system, we carried out the same analysis for other nonlinear complex systems such as the Chua circuit, the R{\"o}ssler system and other autonomous dissipative flows as summarized in section~\ref{Variability}. Figure~\ref{fig:systems} shows the results for their optimal values of $a$ scattered against the standard deviation of the input. In addition, we also constructed combinations of the systems used, in order to fill the gap in between the standard deviations of the Halvorsen model (0.53) and the modified Lorenz system (1.56). We can clearly see that there is a relationship between the optimal $a$ and the input standard deviations. This makes intuitively sense, since the dynamical regime of the hyperbolic tangent needs to be at a different range for different distributions. Surprisingly, this seems to dominate effects of other system-specific properties.. Therefore, as a rule of thumb, the optimal value for the nonlinear scaling parameter is given by $a_{opt} = c / \sigma(\textbf{W}_{in} \textbf{u})^b$ with $b=0.80$ and $c=1.22$ determined by the fitted red curve. This provides a good starting point for the hyperparameter optimization. However, it is always recommended to run a system specific analysis as shown in Fig~\ref{fig:optimalA}. On the example of the Chua circuit it turns out that good predictions cannot only be achieved by values for $a$ close to the result given by the above formula for $a_{opt}$. However, among those systems the Chua circuit yields optimal predictions not only for $a=7.05$ following the above introduced rule of thumb, but also shows another peak for small values of around $a=0.75$ as shown in Fig~\ref{fig:chua}. This might be related to the fact that the equations of the Chua attractor only have a local nonlinearity at $x = \pm 1$, making the linear regime very successful anywhere else. We also looked at a larger parameter range for $a$ and found that the average forecast horizon is monotonically declining for values of $a>10$, which are not shown here. Equivalent results for $a_{opt}$ are gained by carrying out the same analysis for the above mentioned systems based on the reproduction of the correlation dimension. In particular the results for the Chua circuit indicate that there is a significant potential for system specific optimizations.

\section{Conclusions and Outlook}
\label{Conclusion}

In this paper, we used reservoir computing to predict and reconstruct attractors for chaotic systems such as the Lorenz system. While other recurrent neural network based approaches often tend to be a black box, the architecture of reservoir computing is simple enough that a systematic analysis of driving properties for good predictions should be possible. The reason is that the reservoir network itself is static, and therefore predictions are deterministic and depend strongly on output weights once trained. Knowing this, we made alterations to the reservoir network structure by removing nodes and their respective edges based on their weights in the output function. This was motivated by two aims: First, understanding how the prediction quality depends on differential properties of the system. Second, investigating by how much a reservoir computing setup can be reduced while still delivering sufficient prediction performance. We found that removing the nodes associated with the largest 10\% of the output weights improves the replication of the climate of the Lorenz system and reduces variability in prediction quality. This is somewhat counterintuitive, as large weights in the output function suggest a strong influence of the respective node in the aggregation of the (correct) output signal. These findings have to be rather interpreted in the sense that some connections from the nodes with the largest output weights obviously impede the reservoir operations and lead to worse predictions. Further research is needed to unveil the relevance and the impact of connections within the reservoir on the prediction results. Furthermore, it turned out that by applying the node removal framework, the network size can be reduced by more than 30\% at comparable prediction quality and up to 60\% while still delivering reasonable performance. This could be helpful when it comes to hardware implementations of the reservoir, as for example neuromorphic computing \cite{tanaka2019recent}.

Moreover, we varied the scaling of the hyperbolic tangent activation function. We showed that for widely used parametrizations of reservoir computing, a high fraction of the arguments of the activation function is in the saturation regime of the hyperbolic tangent. This leads to high variability and bad prediction quality, as the system cannot adequately grasp the input dynamics. By tuning the scale of the activation function, this problem can be addressed much more conveniently and intuitively than by varying the spectral radius and $W_{in}$ scale separately. We found a relationship between the optimal choice of $a$ and the standard deviation of the input, that can serve as a rule of thumb and provide a good starting point for a complete hyperparameter optimization. At the same time, a system specific analysis and optimization of the nonlinear scaling parameter can unveil interesting results. An example for this was presented for the Chua circuit, where we found not only one peak for the optimal value of $a$ but another -- much smaller -- regime where good predictions can be achieved. We showed that a description of the dependency of the optimal $a$ on the standard deviation of the input of the activation function does not only hold for the Lorenz system but for other complex nonlinear systems as well.\\
Our results demonstrate that a large optimization potential lies in a systematical refinement of the differential reservoir properties for a given data set. This was outlined on the examples of controlled node removal and introduction of a scaling factor in the activation function.
Future research will focus on deepening the understanding of how other differential properties of the reservoir affect the quality of the predictions, with the aim to identify an optimal reservoir in terms of (minimal) size, (best) prediction quality and (highest) statistical robustness.


\section*{Acknowledgements}
We wish to acknowledge useful discussions and comments from Sebastian Baur, Youssef Mabrouk and Mierk Schwabe.

\section*{Data Availability}
The data that support the findings of this study are available from the corresponding author upon reasonable request.

\bibliography{bibliography}

\end{document}